\documentclass[aps,pra,preprint,groupedaddress,showpacs,endfloats*]{revtex4-1}

\usepackage{graphicx}
\usepackage[usenames]{color}

\usepackage{amsmath,amssymb}
\usepackage{hyperref}

\begin{document}

\title{Photoionization Rates of Cs Rydberg Atoms in a 1064 nm Far Off-Resonance Trap}

\author{J. Tallant, D. Booth, and J. P. Shaffer}
\email[]{shaffer@nhn.ou.edu}
\affiliation{Homer L. Dodge Department of Physics and Astronomy, The University of Oklahoma, 440 W. Brooks St. Norman, OK 73019, USA}

\date{\today}

\begin{abstract}
Experimental measurements of photoionization rates of $nD_{5/2}$ Rydberg states of Cs ($50 \leq n \leq 75$) in a 1064 nm far off-resonance dipole trap are presented.  The photoionization rates are obtained by measuring the lifetimes of Rydberg atoms produced inside of a 1064 nm far off-resonance trap and comparing the lifetimes to corresponding control experiments in a magneto-optical trap.  Experimental results for the control experiments agree with recent theoretical predictions for Rydberg state lifetimes and measured photoionization rates are in agreement with transition rates calculated from a model potential.
\end{abstract}


\maketitle

\section{Introduction}

The interest in ultracold Rydberg atoms has sustained for over a decade.  Ultracold Rydberg atoms are readily generated from ultracold samples of ground state atoms inside traps that are relatively straightforward to create.  Experiments with ultracold alkali Rydberg atoms can provide important tests for the theoretical calculation of dipole matrix elements, core-polarizabilities, and lifetimes \cite{Gallagher}.  Ultracold Rydberg atoms have already been used to make precision measurements of quantum defects \cite{QDefect1,Qdefect2}, and radiative lifetimes \cite{RadLife1,RadLife2,RadLife3}.  Recently, ultracold Rydberg atoms have also been used to study exotic states of matter such as long-range diatomic molecules, and triatomic molecules.  Molecular resonances have been observed in Rb and Cs ultracold Rydberg gases \cite{LRResonance,PICollision}, long-range molecules whose bond is stabilized by a dc electric field have been observed \cite{CsRydbergs,Macrodimers,LRWells}, and long-range diatomic and triatomic molecules with a novel binding mechanism have been studied \cite{Trilobite,Trimer}.  High densities are a requirement to investigate these molecules, and far-off resonance traps (FORTs) not only provide high density, but the ground state atoms do not have to be spin-polarized.  However, FORT light can photoionize Rydberg atoms so it is important to measure this effect.  Similarly, photoionization will also play a role in the study of Rydberg atom dipolar quantum gases in high density FORTs.  In addition to these advances, ultracold Rydberg atoms have recently received attention because they can be used for the development of neutral atom quantum gates \cite{QIP1,FORTtheory,QIP2,QIP3}.  For quantum gate schemes such as \cite{QIP2,QIP3}, single qubit operations are performed in a 1064 nm FORT or lattice using high-lying Rydberg states.  The operational performance of multi-qubit gates depends strongly on the lifetime of the constituent Rydberg atoms, and because FORTs are regions of high intensity, photoionization plays a critical role in determining the lifetime of the Rydberg state \cite{FORTtheory}.  In fact, photoionization of Rydberg states serves as a method of detection in these experiments \cite{QIP1,QIP2,FORTtheory}.  In this paper, we systematically investigate photoionization of ultracold Cs $nD_{5/2}$ Rydberg atoms ($50\leq n \leq 75$) in a 1064 nm FORT because of the importance of these measurements for all of these research areas.

To measure the photoionization rates of the Cs Rydberg atoms, their lifetime must be measured inside of the FORT and compared to some control experiment to account for the radiative and blackbody depopulation rates that exist without the FORT light. As a control experiment, we have measured the Cs Rydberg atom lifetimes in a magneto-optical trap (MOT) for each corresponding Rydberg state in the FORT.  The FORT is loaded directly from this MOT so the radiative and blackbody depopulation rates in both experiments are the same.  Only a handful of Rydberg atom lifetime measurements using ultracold atoms are available in the literature \cite{Marcassa1,RadLife1,RadLife2,PIcross,RadLife3,Beterov,CsRyd}, of which only one investigates Cs Rydberg atoms from $30\leq n \leq 40$ \cite{CsRyd}.  The radiative and blackbody depopulation rates measured here are compared to recent theoretical calculations \cite{RydLifes}. While the radiative and blackbody depopulation rates provide important information, there has not been a systematic experimental study of the effect of photoionization on high-lying Rydberg states in a 1064 nm FORT.  Photoionization cross sections of Rb Rydberg atoms ($n=16-20$) in the presence of 10.6 $\mu$m light have been determined \cite{PIcross}, but the atoms were not trapped in the photoionizing field and the light in that case ionizes all nearby states that become populated by blackbody redistribution.  Theoretical photoionization rates in a FORT or lattice have been calculated \cite{FORTtheory,PIFORL} and, where applicable, the two are in agreement.  We calculate photoionization rates according to \cite{FORTtheory,Sobelman} to compare to our experimental measurements.

\section{Experimental Setup}

A Cs vapor-cell MOT is prepared in a stainless steel vacuum chamber.  The background pressure is $\sim$ 5 x $10^{-10}$ torr.  The MOT is formed between two plates used for ramped-field ionization, and is situated 26.5 cm above a z-stack micro-channel plate (MCP) detector \cite{RydTagging}.  The quadrupole magnetic field gradient for the MOT is 12.4 G cm$^{-1}$.  Typical number densities in the MOT are $\sim2$ x $10^{10}$ cm$^{-3}$.  Stray magnetic fields are compensated for by three pairs of orthogonal shim coils such that the viscous damping from optical molasses is maximized.  The MOT is prepared using single-frequency actively-stabilized diode lasers.  The trapping light is produced from a distributed-feedback laser system, which is amplified by a tapered amplifier.  The amplified light is then spatially filtered by a single-mode polarization-preserving fiber.  Repumping light is produced from a separate home-built diode laser system.  The intensity and frequency of all beams creating the MOT are controlled by acousto-optic modulators (AOMs).

\begin{figure}[htpb]
\includegraphics[width=3.0in]{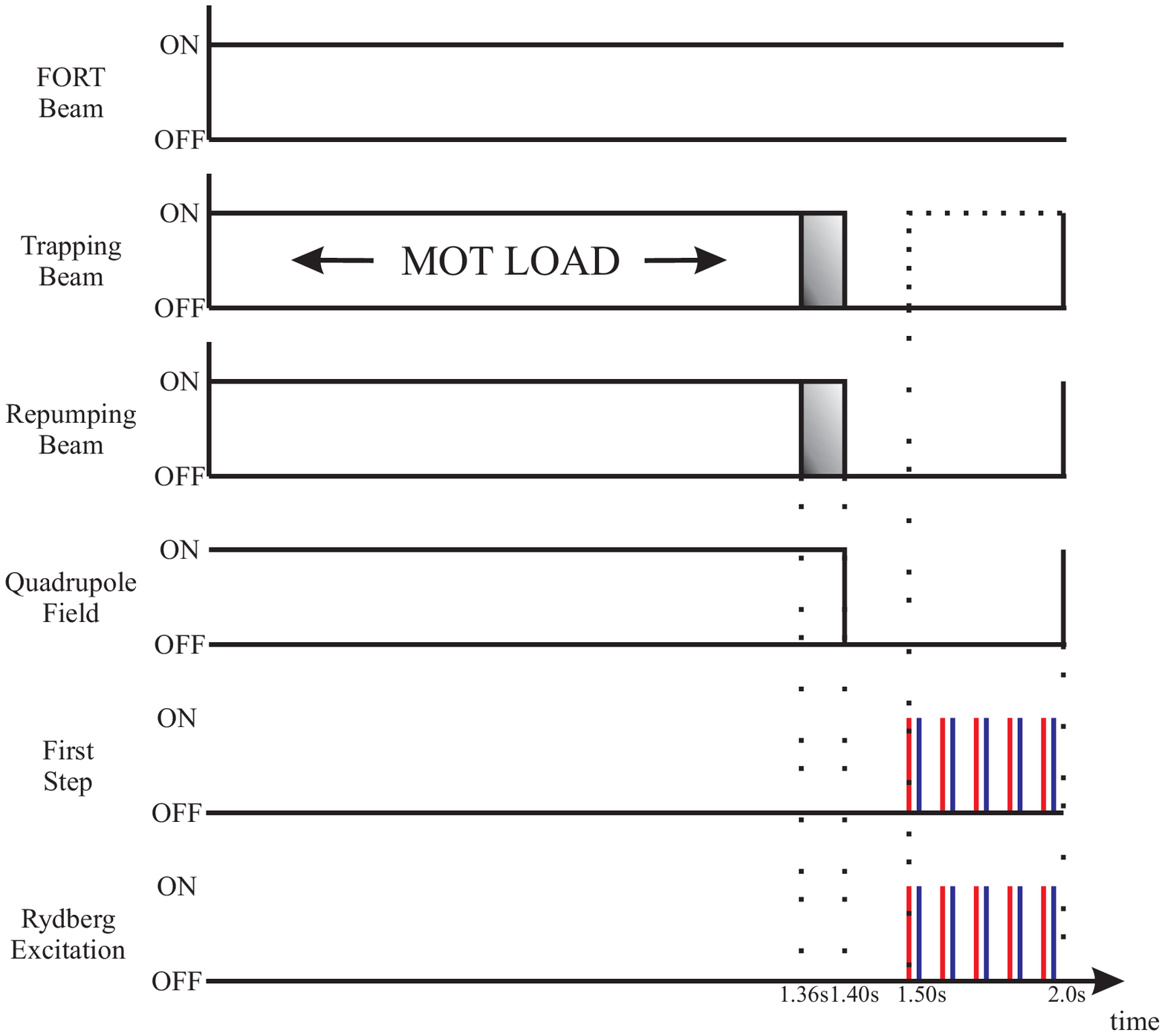}
\caption{(Color Online) Timing for FORT loading and Rydberg atom excitation.  The shaded areas indicate the FORT loading time; the laser beams are on, but their parameters have changed (see text).}
\label{fig:timing}
\end{figure}

A FORT is prepared from a 10 W Yb fiber laser operating at 1064 nm.  The laser beam is focused near the MOT to a spot size ($1/e^{2}$) of 86 $\mu$m $\pm\,1.1$ $\mu$m.  The focused spot size was verified with a CCD camera.  For 7.5 W of power at the MOT location, the FORT depth is $\sim$ 670 $\mu$K.  The radial trap frequency is $\omega_r=2\pi\cdot1.5$ kHz and the axial trap frequency is $\omega_z=2\pi\cdot8.4$ Hz.  The FORT is loaded with the timing sequence illustrated in Fig. \ref{fig:timing}. The MOT is loaded for 1.36 s, with an optimum trapping beam detuning of $\sim -1.5 \Gamma$, where $\Gamma$ is the natural linewidth of the Cs D2 transition ($\Gamma =2\pi \cdot 5.22$ MHz).   The maximum number density in the FORT is 2.5 x $10^{12}$ cm$^{-3}$.
The FORT is loaded for 40 ms.  To load the FORT, the trapping beam is suddenly tuned to the red by $\sim 15$ MHz, while decreasing the intensity by a factor of 3.  Simultaneously, the repumping beam intensity is reduced by a factor of $\sim 100$.  The temperature of the atoms in the FORT is $\sim$ 40 $\mu$K.  Immediately after FORT loading, the quadrupole field for the MOT is turned off within 220 $\mu$s by an insulated-gate bipolar transistor, the MOT laser beams are extinguished and the Cs atoms are trapped inside the FORT.  It should be noted that the quadrupole field may also be turned off immediately before the FORT loading phase, so that the FORT loads directly from an optical molasses.  Due to the low repumping laser intensity and increased detuning of the trapping laser, the atoms are efficiently optically pumped into the $6^{2}S_{1/2}(F=3)$ ground state.  Optically pumping into the absolute ground state eliminates hyperfine-changing collisions within the FORT volume, and leads to an increased trap lifetime \cite{HFchange}.  The measured ($1/e$) lifetime of the FORT loaded from a vapor cell MOT is $\sim800$ ms.

Atoms prepared either in the MOT or in the FORT are excited to high-lying Rydberg states, $50 \leq n \leq 75$, by a two-photon process.  The first step of excitation is an $\sim852$ nm photon and final Rydberg excitation is achieved by absorption of a $\sim509$ nm photon. A Coherent 699-21 ring-dye laser is used to generate linearly-polarized $\sim 509$ nm light, which has a linewidth of $\sim 1.5$ MHz.  The laser beam passes through an AOM before being coupled into a single-mode polarization-preserving fiber.  The fiber output is focused through the trapped atom sample, with a spot size of 58 $\mu$m $\pm\,1.0$ $\mu$m. The two-photon intensity is adjusted so that, on average, one Rydberg atom is excited per laser shot.  The quantum efficiency of the detection system is 0.46 $\pm\,0.06$ \cite{PICollision}.
\begin{figure}[htpb]
\includegraphics[width=3.0in]{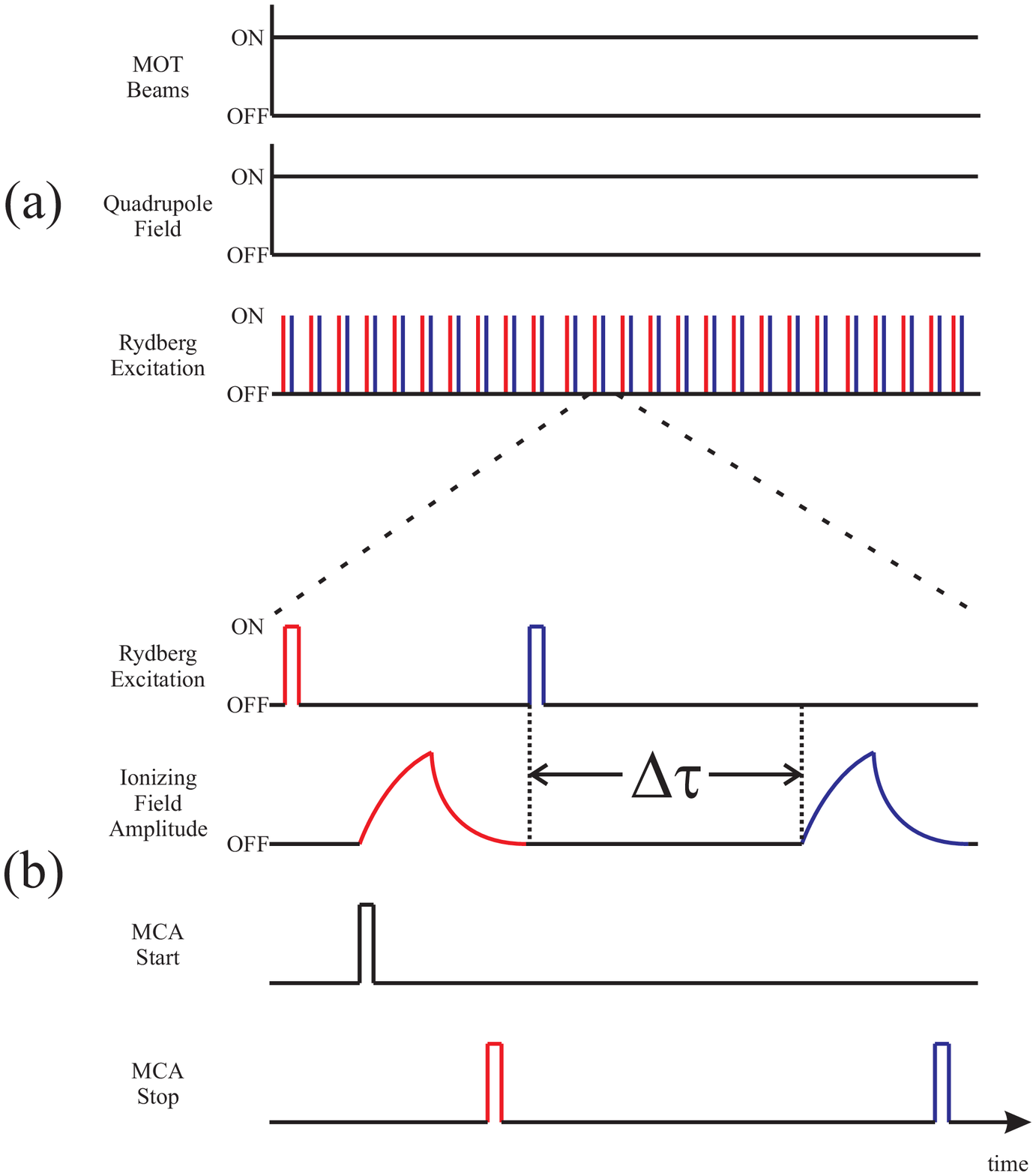}
\caption{(Color Online) (a) Timing for Rydberg atom excitation for control experiment inside the MOT. (b)  Magnified view of Rydberg atom excitation and subsequent ion detection (valid for both control and FORT experiments).  Distributions from the red pulses are used to normalize the distributions acquired from the blue pulses (see text).}
\label{fig:timingdetect}
\end{figure}
After Rydberg excitation, the atoms are ionized by an electric field ramp which is nearly linear in time (Fig. \ref{fig:timingdetect}(b)).  The linear ramp temporally separates different principal quantum number states in the time-of-flight distribution.  The amplitude of the electric field ramp is set just above ionization threshold of the target Rydberg state.  The rise-time of the ionizing ramp is selected to disperse the different Rydberg states for a particular measurement.  Experimental values of the rise-time used ranged from 3.5 $\mu$s -- 7.0 $\mu$s. Once the Rydberg atom is ionized, the positive ion is projected downward onto the MCP detector, which can discriminate ions arriving 500 ps apart.  The ion signal is amplified and sent to a constant-fraction discriminator (CFD).  The output of the CFD is then sent to a multi-channel analyzer (MCA) where a computer records the arrival time distribution of the ion counts.  By adjusting the time between Rydberg atom excitation and ionization, $\Delta\tau$, the Rydberg atom lifetime may be determined from the TOF distributions as described in Sec. \ref{sec:analysis}.

Verifying the lifetimes of Rydberg atoms created inside the MOT region serves as a control for the photoionization experiment in the FORT.  The theoretical Rydberg atom lifetimes for the MOT experiment are calculated according to \cite{RydLifes}, and are described in Sec. \ref{sec:theory}.  Rydberg atom lifetime data are acquired continuously inside the MOT because the MOT is continuously loaded from the background vapor \ref{fig:timingdetect}(a).  Due to imperfect switching, a small amount of Rydberg excitation light leaks through the AOM.  If this effect is not accounted for in some way, the lifetime data become polluted with Rydberg atoms that were unintentionally created at unknown times.  We compensate for this by exciting Rydberg atoms and ramped-field ionizing them with a normalizing pulse immediately before the excitation and ionization sequence that is used for lifetime analysis, see Fig. \ref{fig:timingdetect}.  This limits the amount of leakage light to the time of an individual experimental sequence instead of the much longer time between individual experiments.  Typical experiments last $\sim150$ $\mu$s, whereas the time between experiments is either 2 ms or 4 ms.  The repetition rate for experiments with $n\geq57$ is 500 Hz, but had to be lowered to 250 Hz for $n<57$ due to technical issues with the TOF spectrometer.  The repetition rates for the lifetime measurements inside of the FORT match the rates in the corresponding control experiments, but the total probe time is limited because the FORT is prepared at a rate of 0.5 Hz.  The total probe time for the FORT experiments for each FORT loading sequence is set to 500 ms, which is smaller than the FORT lifetime.

State-dependent photoionization rates are determined by measuring various Rydberg state lifetimes inside of the FORT.  Once atoms are trapped in the FORT, Rydberg atom excitation is delayed by 100 ms to allow atoms previously trapped in the MOT to fall out of the interrogation region.  In order to be promoted to $nD_{5/2}$ Rydberg states, the atom absorbs light which is nearly resonant with the repumping transition required for the MOT.  This $\sim$852 nm light is generated from an independent diode-laser with a linewidth $\sim1$ MHz, which is tuned to compensate for the local (blue) ac Stark shift of the repumping transition.  The light for this ``first step'' laser is sent through an AOM and spatially filtered by a single-mode polarization-preserving fiber.  The output is copropagated with the Rydberg excitation light (509 nm) and both beams are then focused onto the atoms in the FORT. Shallow crossing geometries were also used, but only to check the fidelity of experiments with copropagating excitation lasers.  The spot size of the first step laser at the position of the FORT is 86 $\mu$m $\pm\,0.8$ $\mu$m, which is larger than the 58 $\mu$m $\pm\,1.0$ $\mu$m excitation beam.  The intensity of the first step laser (852 nm) is set to avoid any radiation pressure effects on the trapped atoms.  The FORT beam and Rydberg excitation beams intersect at an angle of $\sim$ 112.5$^{\circ}$.

\section{Theoretical Lifetimes\label{sec:theory}}
In the absence of the FORT beam, the total decay rate out of the Rydberg state is the sum of the radiative decay rate, $\gamma_{rad}$, and the blackbody decay rate, $\gamma_{bb}$.  To calculate the effective lifetimes of the Rydberg atoms in the absence of the FORT beam, we used a model developed by Beterov \cite{RydLifes}.  This model is accurate to within 5\% of the lifetime for $15<n<80$.  The lifetime of atoms prepared in the MOT, $\tau_{MOT}$, can be expressed in terms of the radiative and blackbody decay rates:

\begin{equation}
\label{eq:motlifetime}
\tau_{MOT} = \left( \gamma_{rad}+\gamma_{bb} \right)^{-1}  .
\end{equation}

Inside of the FORT, the intense field of the trapping laser causes significant photoionization of Rydberg atoms, which leads to a decreased Rydberg atom lifetime inside the FORT beam.  To calculate the effect of photoionization on the Rydberg atom lifetime, we start by calculating the photoionization cross section, $\sigma_{PI}$, according to \cite{Gallagher},

\begin{equation}
\label{eq:PICrossSection}
\sigma_{PI} = \left. 2 \pi^2 \frac{\hbar e^2}{m_e c} \frac{df}{dE} \right|_{E = E_{r} + \hbar \omega}.
\end{equation}
$E_r$ is the energy of the Rydberg state and $\hbar\omega$ is the photon energy of the trapping laser.  The oscillator strength distribution for coupling to the continuum, $df/dE$, can be written as \cite{Gallagher,Sobelman}

\begin{equation}
\label{eq:oscstr}
\frac{df}{dE} = \sum_{L=L_r - 1}^{L_r +1} \frac{2 m_e \omega L_{>}}{3 \hbar \left(2 L_r + 1\right)} \left|\int \psi_{n,l}(r) r \phi_{L,E}(r) dr \right|^2.
\end{equation}
$L_r$ is the orbital angular momentum quantum number of the Rydberg state and $L_>$ is the greater of $L$ and $L_r$.  The wavefunction of the Rydberg state is $\psi_{n,l}(r)$, and $\phi_{L,E}(r)$ is the continuum wavefunction of energy, $E$.  To verify the theoretically calculated photoionization cross sections, we reproduced the values listed in \cite{FORTtheory} for the 50$D$, 70$D$, and 90$D$ states of Rb.

The bound and continuum wavefunctions are calculated numerically with RADIAL \cite{Radial}, using the $l$-dependent potential \cite{DispersionCoefs}.  The continuum wavefunctions are normalized per unit energy \cite{FORTtheory},

\begin{equation}
\label{eq:continuumnormalization}
\phi_{L,E}(r) = \sqrt{\frac{2 m_e}{\pi \hbar^2 k}} \Phi_{L,E}(r),
\end{equation}
$\Phi_{L,E}(r)$ is the normalized continuum wavefunction from RADIAL and $k=\hbar^{-1}\sqrt{2m_eE}$.

Using the photoionization cross section from Eq. \ref{eq:PICrossSection}, the average photoionization rate, $\gamma_{PI}$, is given by,

\begin{equation}
\label{eq:PIrate}
\gamma_{PI} = \frac{I}{\hbar\omega}\,\sigma_{PI}.
\end{equation}
The average intensity of the trapping laser over the excitation region is $I$.  Since the intensity of the laser is a function of the distance from the center of the beam, atoms at different locations in the trap will have different photoionization rates.  A Gaussian intensity distribution for the trapping light was used to compute the average intensity.  Inside of the FORT, the photoionization rate adds to the existing decay rates so that the reduced Rydberg atom lifetime is simply expressed by,

\begin{equation}
\label{eq:RydLifetime}
\tau_{Ry}=\left(\gamma_{rad}+\gamma_{bb}+\gamma_{PI}\right)^{-1}.
\end{equation}
By determining $\gamma_{rad}$ and $\gamma_{bb}$ for a given Rydberg state in the MOT control experiment, $\gamma_{PI}$ may be determined by measuring the lifetime of the Rydberg atoms trapped inside of the FORT.  $\gamma_{bb}$ is the same in the MOT and FORT experiments because the local environment is the same in both measurements.

\section{Results and Analysis\label{sec:analysis}}
The $nD$ Rydberg atoms excited in this work are low-field seekers.  The trapping potential in the FORT that exists for the ground state atoms becomes a repulsive potential once the Rydberg atom is excited.  To estimate the magnitude of this repulsive energy, we assume the polarizability of the Rydberg states correspond to the polarizability of a free electron, $-e^2/m_e\omega^2$, where $\omega$ is the angular frequency of the trapping beam.  This overestimates the repulsive effect.  Using this estimate, the $nD$ states would be expelled from the FORT in $\sim1$ ms if they traveled directly along the radial dimension of the FORT.  This has the largest effect for the 75$D_{5/2}$ state, which has both the highest polarizability and longest lifetime.  Because the Rydberg atoms must accelerate along the gradient of intensity out of the trapping volume, the force on Rydberg atoms in the center is initially very small.  The repulsive force moves the Rydberg atoms $<1$ $\mu$m in 500 $\mu$s.  Thus, for times $\lesssim1$ ms, the Rydberg atom motion is dominated by the initial temperature.  A Cs Rydberg atom at $\sim40$ $\mu$K moves $\sim25$ $\mu$m in 500 $\mu$s.  These time scales allow for adequate determination of the photoionization rate for the $75D_{5/2}$ state in the average intensity of the trapping field.  Photoionization rates for higher-lying Rydberg states, however, would have to take into account the Rydberg atom motion to regions of lower intensity.

To determine the Rydberg atom lifetimes, TOF distributions are accumulated for different delay times, $\Delta\tau$, between Rydberg atom excitation and ramped-field ionization.  A TOF distribution from a single delay contains a distribution of counts from the normalizing pulse (red pulses in Fig. \ref{fig:timingdetect}) and a temporally separated ($\gtrsim100$ $\mu$s) distribution of counts used for analysis.  For the earliest delays used ($\Delta\tau=5$ $\mu$s), a prominent Rydberg atom peak appears in the TOF distribution that is $\sim10$ $\mu$s wide, depending on the ionizing field amplitude used.  At later delay times peaks from higher principal quantum number states begin to appear $\sim15$ $\mu$s before the target Rydberg state in the TOF distribution.  These peaks are due to blackbody redistribution and are $\sim10$ $\mu$s wide, allowing easy discrimination from the target Rydberg atom peak.

\begin{figure}[htpb]
\includegraphics[width=3.0in]{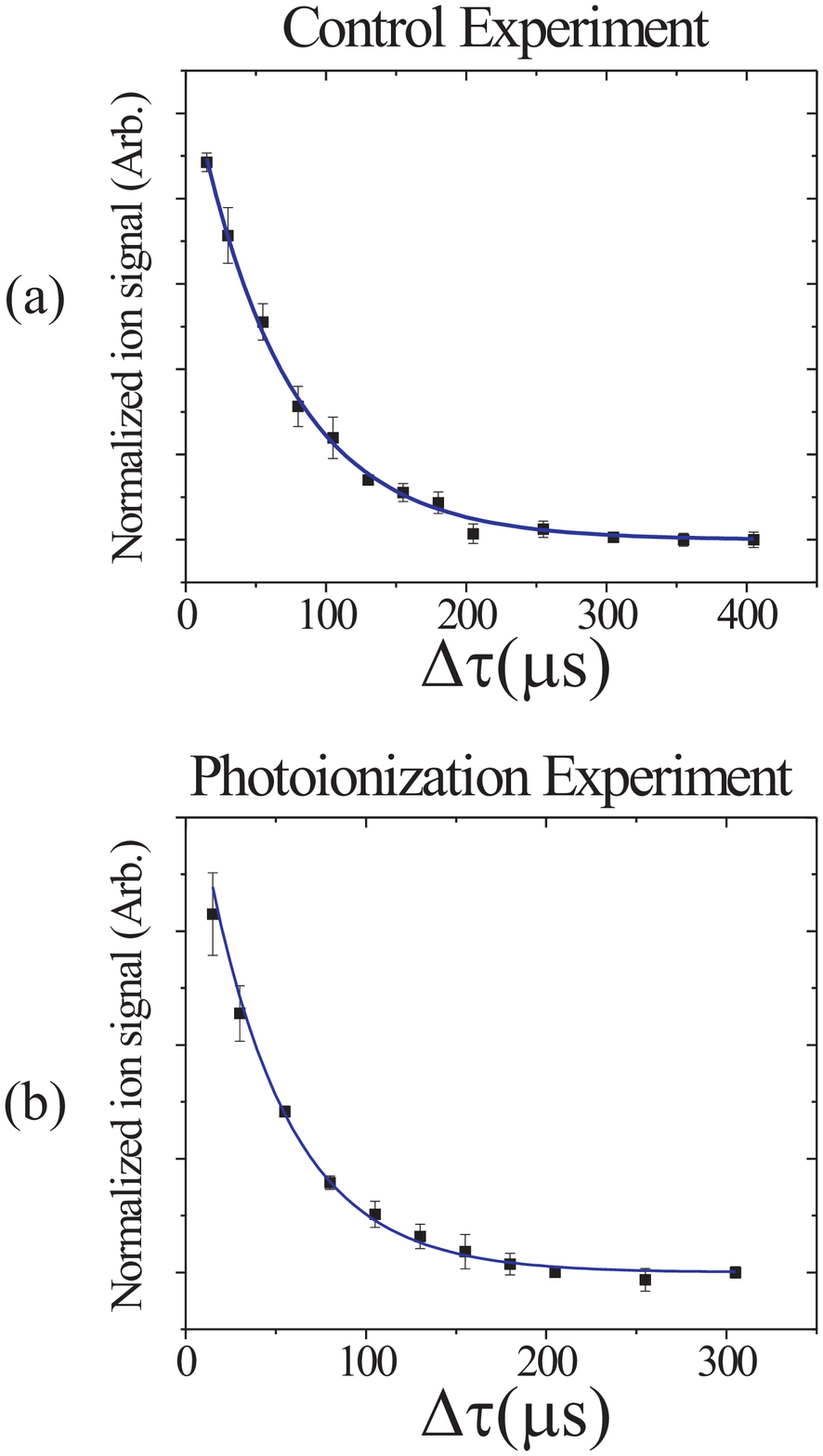}
\caption{(Color Online) Experimental lifetime data of the $55D_{5/2}$ Rydberg state.  Error bars are the standard deviation of three measurements. (a)  Rydberg atom lifetime measurement in the MOT.  The solid blue line is a fit to the data with $\tau_{Ry}=65\pm3$ $\mu$s. (b)  Rydberg atom lifetime data inside of the FORT.  The solid line is a fit to the data with $\tau_{Ry}=45\pm3$ $\mu$s.}
\label{fig:55DLifes}
\end{figure}

Rydberg atom lifetime analysis is carried out the same way for the control experiment and the photoionization experiment.  An integration window is centered on the target Rydberg atom peak that is 1-5 $\mu$s wide, depending on the ramped-field amplitude.  The number of ion counts falling inside the window are normalized by the total number of counts appearing in the normalization distribution arriving $\gtrsim100$ $\mu$s earlier.  The normalization tends to cancel effects from excitation laser fluctuations in intensity or frequency as well as trapped atom number fluctuations between experiments.  The normalized ion counts are then plotted as a function of the delay between Rydberg atom creation and ionization, $\Delta\tau$.  Experimental data for the MOT control measurement at $55D_{5/2}$ are shown in Fig. \ref{fig:55DLifes}(a).
The data are fit to a decaying exponential in time, $Ae^{-\Delta\tau/\tau_{Ry}}$.  The amplitude of the decay is $A$, $\Delta\tau$ is the delay time between excitation and ionization, and $\tau_{Ry}$ is the $1/e$ lifetime of the target Rydberg state.  The amplitude of the decay and the Rydberg atom lifetime are the only free parameters.  The lifetime of the $55D_{5/2}$ state in the MOT control experiment is measured to be $\tau_{Ry}=65\pm3$ $\mu$s.  Theoretical calculations for this state predict the lifetime is $\sim63$ $\mu$s \cite{RydLifes}.  Data for the FORT photoionization experiment on the $55D_{5/2}$ state are shown in Fig. \ref{fig:55DLifes}(b).  The enhanced photoionization rate from the FORT beam has a significant effect on the Rydberg atom lifetime.  The measured lifetime has decreased to $\tau_{Ry}=45\pm3$ $\mu$s, in agreement with theoretical predictions from \cite{FORTtheory}.  The corresponding photoionization rate is 4.9 kHz $\pm\,0.6$ kHz.

\begin{figure}[htpb]
\includegraphics[width=3.0in]{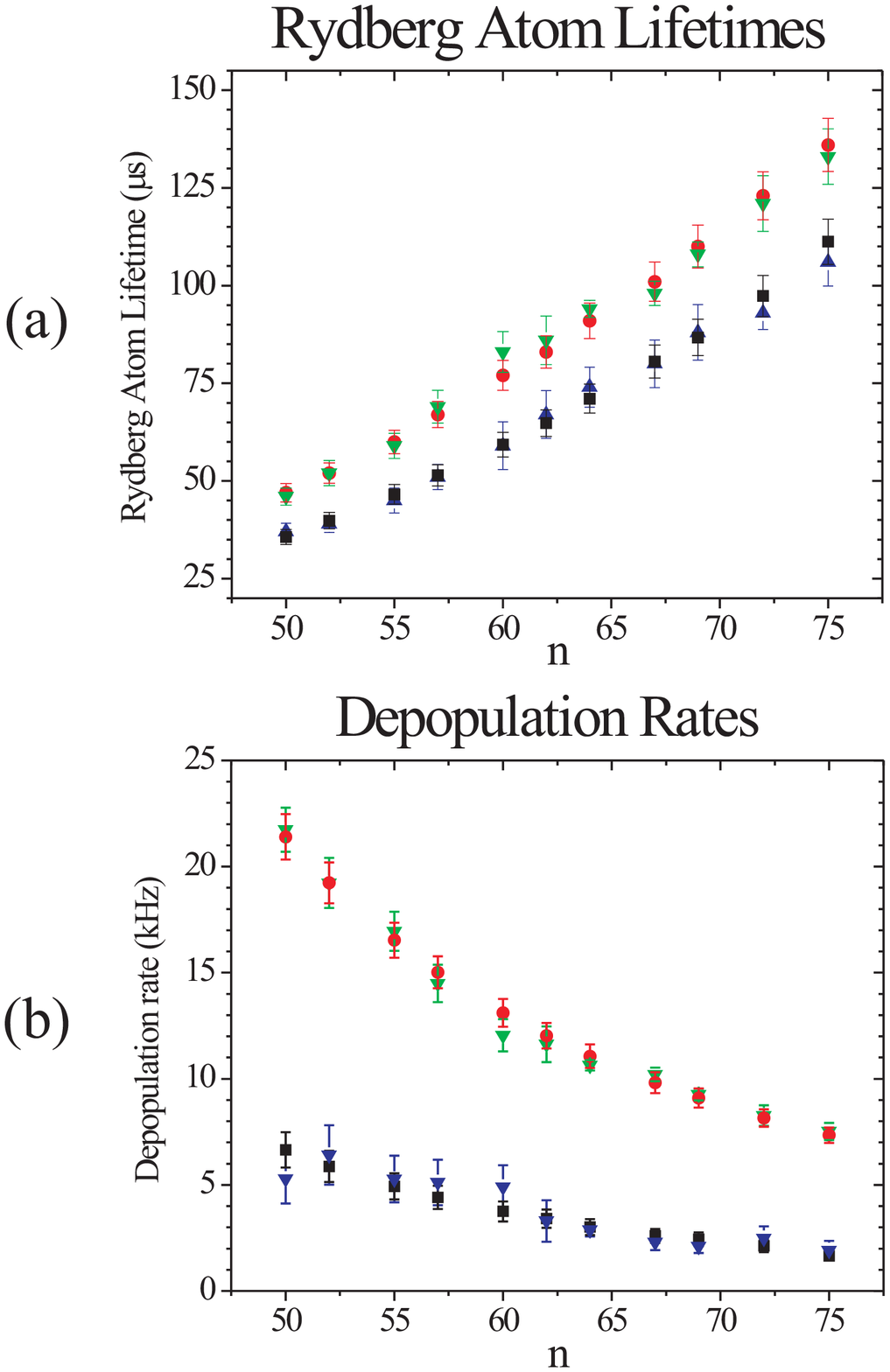}
\caption{(Color Online) (a) Experimental and theoretical Rydberg atom lifetimes as a function of principal quantum number. \textcolor{green}{$\blacktriangledown$} Experimental control lifetimes.  \textcolor{red}{$\bullet$} Theoretical lifetime for the control experiment. \textcolor{blue}{$\blacktriangle$} Experimental lifetimes inside the FORT.  $\blacksquare$ Theoretical lifetime for 129 kW cm$^{-2}$ of 1064 nm light.  (b) Experimental and theoretical depopulation rates as a function of principal quantum number. \textcolor{green}{$\blacktriangledown$} Total experimental decay rate in control measurement.  \textcolor{red}{$\bullet$} Theoretical decay rate for the control experiment.  \textcolor{blue}{$\blacktriangle$} Experimental photoionization rate.  $\blacksquare$ Theoretical photoionization rate for 129 kW cm$^{-2}$ of 1064 nm light \cite{FORTtheory}.  Error bars for the photoionization rates include a 17 kW cm$^{-2}$ uncertainty in the intensity at the trap location.}
\label{fig:CombinedData}
\end{figure}

Lifetimes were measured for 11 $nD_{5/2}$ Rydberg states, $50\leq n \leq 75$, produced in a 1064 nm FORT and compared to the corresponding MOT control experiments.  Experimental data for the Rydberg state lifetimes are compared to the theoretical values for both the MOT control and FORT photoionization measurements in Fig. \ref{fig:CombinedData}(a).  The error in the theoretical calculations for the control lifetimes is $\pm5\%$ \cite{RydLifes}.   Experimental error bars come from a convolution of the error in the exponential fit and the 1 $\mu$s excitation time.

The theoretical and experimental depopulation rates are shown in Fig. \ref{fig:CombinedData}(b).  The sum of the radiative and blackbody decay rates is determined from the control measurement.  The total decay rate from the control experiment is subtracted from the total decay rate in the FORT to yield the measured photoionization rate, $\gamma_{PI}$.  The error in the photoionization rate is generated from an estimated uncertainty in the trapping beam intensity of 17 kW cm$^{-2}$.  The uncertainty corresponds to the measured uncertainty in the waist radius of 43 $\mu$m $\pm\,0.6$ $\mu$m, and an estimated power at the trap location of 7.5 W $\pm\,0.2$ W.

The experimental lifetimes in the control experiment agree well with the predictions from \cite{RydLifes}, which suggests that we are also in agreement with experimental results for Rb \cite{RadLife1,RadLife2,RadLife3,Beterov} and lower-lying states of Cs \cite{CsRyd}.  Consistency with theoretical calculations and experimental work of other groups gives us confidence that the depopulation rates obtained in the MOT control experiment are valid and may safely be used to obtain photoionization rates in the 1064 nm FORT experiment.  The experimental photoionization rates also agree well with our calculations following \cite{FORTtheory}.  Over most of the range of validity of the model by Beterov \cite{RydLifes}, our experimental results show that photoionization rates from a FORT can be generated from principles following \cite{FORTtheory,Sobelman}.

\section{Conclusions}
We have measured Rydberg atom lifetimes for $nD_{5/2}$ states of Cs, where ($50 \leq n \leq 75$).  Measurements took place inside of a MOT to obtain the total depopulation rates due to radiative decay and blackbody redistribution.  These depopulation rates were subtracted from the total depopulation rates in the corresponding lifetime experiments inside the FORT to obtain experimental photoionization rates for the Rydberg states.  The FORT and MOT measurements are carried out at nearly identical positions in the experimental apparatus.  The depopulation rates due to spontaneous emission and blackbody radiation are in agreement with recent calculations \cite{RydLifes} and are thus in agreement with experimental results for lower-lying states of Rb \cite{RadLife1,RadLife2,RadLife3,Beterov} and Cs \cite{CsRyd}.  There is no systematic photoionization experiment with which to compare to our data, but the photoionization rates observed here are consistent with calculations following \cite{FORTtheory,Sobelman}.  This work provides experimental evidence that these methods can be safely used to predict photoionization rates in a far off-resonance trap.

\begin{acknowledgments}
The authors would like to thank L. G. Marcassa for enlightening discussions.  This work is supported by the National Science Foundation (PHY-0855324) and the Army Research Office (W911NF-08-1-0257).
\end{acknowledgments}


\begin{thebibliography}{28}%
\makeatletter
\providecommand \@ifxundefined [1]{%
 \@ifx{#1\undefined}
}%
\providecommand \@ifnum [1]{%
 \ifnum #1\expandafter \@firstoftwo
 \else \expandafter \@secondoftwo
 \fi
}%
\providecommand \@ifx [1]{%
 \ifx #1\expandafter \@firstoftwo
 \else \expandafter \@secondoftwo
 \fi
}%
\providecommand \natexlab [1]{#1}%
\providecommand \enquote  [1]{``#1''}%
\providecommand \bibnamefont  [1]{#1}%
\providecommand \bibfnamefont [1]{#1}%
\providecommand \citenamefont [1]{#1}%
\providecommand \href@noop [0]{\@secondoftwo}%
\providecommand \href [0]{\begingroup \@sanitize@url \@href}%
\providecommand \@href[1]{\@@startlink{#1}\@@href}%
\providecommand \@@href[1]{\endgroup#1\@@endlink}%
\providecommand \@sanitize@url [0]{\catcode `\\12\catcode `\$12\catcode
  `\&12\catcode `\#12\catcode `\^12\catcode `\_12\catcode `\%12\relax}%
\providecommand \@@startlink[1]{}%
\providecommand \@@endlink[0]{}%
\providecommand \url  [0]{\begingroup\@sanitize@url \@url }%
\providecommand \@url [1]{\endgroup\@href {#1}{\urlprefix }}%
\providecommand \urlprefix  [0]{URL }%
\providecommand \Eprint [0]{\href }%
\providecommand \doibase [0]{http://dx.doi.org/}%
\providecommand \selectlanguage [0]{\@gobble}%
\providecommand \bibinfo  [0]{\@secondoftwo}%
\providecommand \bibfield  [0]{\@secondoftwo}%
\providecommand \translation [1]{[#1]}%
\providecommand \BibitemOpen [0]{}%
\providecommand \bibitemStop [0]{}%
\providecommand \bibitemNoStop [0]{.\EOS\space}%
\providecommand \EOS [0]{\spacefactor3000\relax}%
\providecommand \BibitemShut  [1]{\csname bibitem#1\endcsname}%
\let\auto@bib@innerbib\@empty
\bibitem [{\citenamefont {Gallagher}(1994)}]{Gallagher}%
  \BibitemOpen
  \bibfield  {author} {\bibinfo {author} {\bibfnamefont {T.~F.}\ \bibnamefont
  {Gallagher}},\ }\href@noop {} {\emph {\bibinfo {title} {Rydberg Atoms}}},\
  \bibinfo {edition} {1st}\ ed.\ (\bibinfo  {publisher} {Cambridge University
  Press},\ \bibinfo {year} {1994})\BibitemShut {NoStop}%
\bibitem [{\citenamefont {Li}\ \emph {et~al.}(2003)\citenamefont {Li},
  \citenamefont {Mourachko}, \citenamefont {Noel},\ and\ \citenamefont
  {Gallagher}}]{QDefect1}%
  \BibitemOpen
  \bibfield  {author} {\bibinfo {author} {\bibfnamefont {W.}~\bibnamefont
  {Li}}, \bibinfo {author} {\bibfnamefont {I.}~\bibnamefont {Mourachko}},
  \bibinfo {author} {\bibfnamefont {M.~W.}\ \bibnamefont {Noel}}, \ and\
  \bibinfo {author} {\bibfnamefont {T.~F.}\ \bibnamefont {Gallagher}},\
  }\href@noop {} {\bibfield  {journal} {\bibinfo  {journal} {Phys. Rev. A}\
  }\textbf {\bibinfo {volume} {67}},\ \bibinfo {pages} {052502} (\bibinfo
  {year} {2003})}\BibitemShut {NoStop}%
\bibitem [{\citenamefont {Han}\ \emph {et~al.}(2006)\citenamefont {Han},
  \citenamefont {Jamil}, \citenamefont {Norum}, \citenamefont {Tanner},\ and\
  \citenamefont {Gallagher}}]{Qdefect2}%
  \BibitemOpen
  \bibfield  {author} {\bibinfo {author} {\bibfnamefont {J.}~\bibnamefont
  {Han}}, \bibinfo {author} {\bibfnamefont {Y.}~\bibnamefont {Jamil}}, \bibinfo
  {author} {\bibfnamefont {D.~V.~L.}\ \bibnamefont {Norum}}, \bibinfo {author}
  {\bibfnamefont {P.~J.}\ \bibnamefont {Tanner}}, \ and\ \bibinfo {author}
  {\bibfnamefont {T.~F.}\ \bibnamefont {Gallagher}},\ }\href@noop {} {\bibfield
   {journal} {\bibinfo  {journal} {Phys. Rev. A}\ }\textbf {\bibinfo {volume}
  {74}},\ \bibinfo {pages} {054502} (\bibinfo {year} {2006})}\BibitemShut
  {NoStop}%
\bibitem [{\citenamefont {de~Oliveira}\ \emph {et~al.}(2002)\citenamefont
  {de~Oliveira}, \citenamefont {Mancini}, \citenamefont {Bagnato},\ and\
  \citenamefont {Marcassa}}]{RadLife1}%
  \BibitemOpen
  \bibfield  {author} {\bibinfo {author} {\bibfnamefont {A.~L.}\ \bibnamefont
  {de~Oliveira}}, \bibinfo {author} {\bibfnamefont {M.~W.}\ \bibnamefont
  {Mancini}}, \bibinfo {author} {\bibfnamefont {V.~S.}\ \bibnamefont
  {Bagnato}}, \ and\ \bibinfo {author} {\bibfnamefont {L.~G.}\ \bibnamefont
  {Marcassa}},\ }\href@noop {} {\bibfield  {journal} {\bibinfo  {journal}
  {Phys. Rev. A}\ }\textbf {\bibinfo {volume} {65}},\ \bibinfo {pages}
  {031401(R)} (\bibinfo {year} {2002})}\BibitemShut {NoStop}%
\bibitem [{\citenamefont {Nascimento}\ \emph {et~al.}(2006)\citenamefont
  {Nascimento}, \citenamefont {Caliri}, \citenamefont {de~Oliveira},
  \citenamefont {Bagnato},\ and\ \citenamefont {Marcassa}}]{RadLife2}%
  \BibitemOpen
  \bibfield  {author} {\bibinfo {author} {\bibfnamefont {V.~A.}\ \bibnamefont
  {Nascimento}}, \bibinfo {author} {\bibfnamefont {L.~L.}\ \bibnamefont
  {Caliri}}, \bibinfo {author} {\bibfnamefont {A.~L.}\ \bibnamefont
  {de~Oliveira}}, \bibinfo {author} {\bibfnamefont {V.~S.}\ \bibnamefont
  {Bagnato}}, \ and\ \bibinfo {author} {\bibfnamefont {L.~G.}\ \bibnamefont
  {Marcassa}},\ }\href@noop {} {\bibfield  {journal} {\bibinfo  {journal}
  {Phys. Rev. A}\ }\textbf {\bibinfo {volume} {74}},\ \bibinfo {pages} {054501}
  (\bibinfo {year} {2006})}\BibitemShut {NoStop}%
\bibitem [{\citenamefont {Branden}\ \emph {et~al.}(2010)\citenamefont
  {Branden}, \citenamefont {Juhasz}, \citenamefont {Mahlokozera}, \citenamefont
  {Vesa}, \citenamefont {Wilson}, \citenamefont {Zheng}, \citenamefont
  {Kortyna},\ and\ \citenamefont {Tate}}]{RadLife3}%
  \BibitemOpen
  \bibfield  {author} {\bibinfo {author} {\bibfnamefont {D.~B.}\ \bibnamefont
  {Branden}}, \bibinfo {author} {\bibfnamefont {T.}~\bibnamefont {Juhasz}},
  \bibinfo {author} {\bibfnamefont {T.}~\bibnamefont {Mahlokozera}}, \bibinfo
  {author} {\bibfnamefont {C.}~\bibnamefont {Vesa}}, \bibinfo {author}
  {\bibfnamefont {R.~O.}\ \bibnamefont {Wilson}}, \bibinfo {author}
  {\bibfnamefont {M.}~\bibnamefont {Zheng}}, \bibinfo {author} {\bibfnamefont
  {A.}~\bibnamefont {Kortyna}}, \ and\ \bibinfo {author} {\bibfnamefont
  {D.~A.}\ \bibnamefont {Tate}},\ }\href@noop {} {\bibfield  {journal}
  {\bibinfo  {journal} {J. Phys. B}\ }\textbf {\bibinfo {volume} {43}},\
  \bibinfo {pages} {015002} (\bibinfo {year} {2010})}\BibitemShut {NoStop}%
\bibitem [{\citenamefont {Farooqi}\ \emph {et~al.}(2003)\citenamefont
  {Farooqi}, \citenamefont {Tong}, \citenamefont {Krishnan}, \citenamefont
  {Stanojevic}, \citenamefont {Zhang}, \citenamefont {Ensher}, \citenamefont
  {Estrin}, \citenamefont {Boisseau}, \citenamefont {C\^{o}t\'{e}},
  \citenamefont {Eyler},\ and\ \citenamefont {Gould}}]{LRResonance}%
  \BibitemOpen
  \bibfield  {author} {\bibinfo {author} {\bibfnamefont {S.~M.}\ \bibnamefont
  {Farooqi}}, \bibinfo {author} {\bibfnamefont {D.}~\bibnamefont {Tong}},
  \bibinfo {author} {\bibfnamefont {S.}~\bibnamefont {Krishnan}}, \bibinfo
  {author} {\bibfnamefont {J.}~\bibnamefont {Stanojevic}}, \bibinfo {author}
  {\bibfnamefont {Y.~P.}\ \bibnamefont {Zhang}}, \bibinfo {author}
  {\bibfnamefont {J.~R.}\ \bibnamefont {Ensher}}, \bibinfo {author}
  {\bibfnamefont {A.~S.}\ \bibnamefont {Estrin}}, \bibinfo {author}
  {\bibfnamefont {C.}~\bibnamefont {Boisseau}}, \bibinfo {author}
  {\bibfnamefont {R.}~\bibnamefont {C\^{o}t\'{e}}}, \bibinfo {author}
  {\bibfnamefont {E.~E.}\ \bibnamefont {Eyler}}, \ and\ \bibinfo {author}
  {\bibfnamefont {P.~L.}\ \bibnamefont {Gould}},\ }\href@noop {} {\bibfield
  {journal} {\bibinfo  {journal} {Phys. Rev. Lett.}\ }\textbf {\bibinfo
  {volume} {91}},\ \bibinfo {pages} {183002} (\bibinfo {year}
  {2003})}\BibitemShut {NoStop}%
\bibitem [{\citenamefont {Overstreet}\ \emph {et~al.}(2007)\citenamefont
  {Overstreet}, \citenamefont {Schwettmann}, \citenamefont {Tallant},\ and\
  \citenamefont {Shaffer}}]{PICollision}%
  \BibitemOpen
  \bibfield  {author} {\bibinfo {author} {\bibfnamefont {K.~R.}\ \bibnamefont
  {Overstreet}}, \bibinfo {author} {\bibfnamefont {A.}~\bibnamefont
  {Schwettmann}}, \bibinfo {author} {\bibfnamefont {J.}~\bibnamefont
  {Tallant}}, \ and\ \bibinfo {author} {\bibfnamefont {J.~P.}\ \bibnamefont
  {Shaffer}},\ }\href@noop {} {\bibfield  {journal} {\bibinfo  {journal} {Phys.
  Rev. A}\ }\textbf {\bibinfo {volume} {76}},\ \bibinfo {pages} {011403(R)}
  (\bibinfo {year} {2007})}\BibitemShut {NoStop}%
\bibitem [{\citenamefont {Schwettmann}\ \emph {et~al.}(2006)\citenamefont
  {Schwettmann}, \citenamefont {Crawford}, \citenamefont {Overstreet},\ and\
  \citenamefont {Shaffer}}]{CsRydbergs}%
  \BibitemOpen
  \bibfield  {author} {\bibinfo {author} {\bibfnamefont {A.}~\bibnamefont
  {Schwettmann}}, \bibinfo {author} {\bibfnamefont {J.}~\bibnamefont
  {Crawford}}, \bibinfo {author} {\bibfnamefont {K.~R.}\ \bibnamefont
  {Overstreet}}, \ and\ \bibinfo {author} {\bibfnamefont {J.~P.}\ \bibnamefont
  {Shaffer}},\ }\href@noop {} {\bibfield  {journal} {\bibinfo  {journal} {Phys.
  Rev. A}\ }\textbf {\bibinfo {volume} {74}},\ \bibinfo {pages} {020701(R)}
  (\bibinfo {year} {2006})}\BibitemShut {NoStop}%
\bibitem [{\citenamefont {Overstreet}\ \emph {et~al.}(2009)\citenamefont
  {Overstreet}, \citenamefont {Schwettmann}, \citenamefont {Tallant},
  \citenamefont {Booth},\ and\ \citenamefont {Shaffer}}]{Macrodimers}%
  \BibitemOpen
  \bibfield  {author} {\bibinfo {author} {\bibfnamefont {K.~R.}\ \bibnamefont
  {Overstreet}}, \bibinfo {author} {\bibfnamefont {A.}~\bibnamefont
  {Schwettmann}}, \bibinfo {author} {\bibfnamefont {J.}~\bibnamefont
  {Tallant}}, \bibinfo {author} {\bibfnamefont {D.}~\bibnamefont {Booth}}, \
  and\ \bibinfo {author} {\bibfnamefont {J.~P.}\ \bibnamefont {Shaffer}},\
  }\href@noop {} {\bibfield  {journal} {\bibinfo  {journal} {Nat. Phys.}\
  }\textbf {\bibinfo {volume} {5}},\ \bibinfo {pages} {581} (\bibinfo {year}
  {2009})}\BibitemShut {NoStop}%
\bibitem [{\citenamefont {Schwettmann}\ \emph {et~al.}(2007)\citenamefont
  {Schwettmann}, \citenamefont {Overstreet}, \citenamefont {Tallant},\ and\
  \citenamefont {Shaffer}}]{LRWells}%
  \BibitemOpen
  \bibfield  {author} {\bibinfo {author} {\bibfnamefont {A.}~\bibnamefont
  {Schwettmann}}, \bibinfo {author} {\bibfnamefont {K.~R.}\ \bibnamefont
  {Overstreet}}, \bibinfo {author} {\bibfnamefont {J.}~\bibnamefont {Tallant}},
  \ and\ \bibinfo {author} {\bibfnamefont {J.~P.}\ \bibnamefont {Shaffer}},\
  }\href@noop {} {\bibfield  {journal} {\bibinfo  {journal} {J. Mod. Opt.}\
  }\textbf {\bibinfo {volume} {54}},\ \bibinfo {pages} {2551} (\bibinfo {year}
  {2007})}\BibitemShut {NoStop}%
\bibitem [{\citenamefont {Bendowsky}\ \emph {et~al.}(2009)\citenamefont
  {Bendowsky}, \citenamefont {Butscher}, \citenamefont {Nipper}, \citenamefont
  {Shaffer}, \citenamefont {Low},\ and\ \citenamefont {Pfau}}]{Trilobite}%
  \BibitemOpen
  \bibfield  {author} {\bibinfo {author} {\bibfnamefont {V.}~\bibnamefont
  {Bendowsky}}, \bibinfo {author} {\bibfnamefont {B.}~\bibnamefont {Butscher}},
  \bibinfo {author} {\bibfnamefont {J.}~\bibnamefont {Nipper}}, \bibinfo
  {author} {\bibfnamefont {J.~P.}\ \bibnamefont {Shaffer}}, \bibinfo {author}
  {\bibfnamefont {R.}~\bibnamefont {Low}}, \ and\ \bibinfo {author}
  {\bibfnamefont {T.}~\bibnamefont {Pfau}},\ }\href@noop {} {\bibfield
  {journal} {\bibinfo  {journal} {Nature}\ }\textbf {\bibinfo {volume} {458}},\
  \bibinfo {pages} {1005} (\bibinfo {year} {2009})}\BibitemShut {NoStop}%
\bibitem [{\citenamefont {Bendkowsky}\ \emph {et~al.}(tion)\citenamefont
  {Bendkowsky}, \citenamefont {Butscher}, \citenamefont {Nipper}, \citenamefont
  {Balewski}, \citenamefont {Shaffer}, \citenamefont {Low}, \citenamefont
  {Pfau}, \citenamefont {Li}, \citenamefont {Stanojevic}, \citenamefont
  {Pohl},\ and\ \citenamefont {Rost}}]{Trimer}%
  \BibitemOpen
  \bibfield  {author} {\bibinfo {author} {\bibfnamefont {V.}~\bibnamefont
  {Bendkowsky}}, \bibinfo {author} {\bibfnamefont {B.}~\bibnamefont
  {Butscher}}, \bibinfo {author} {\bibfnamefont {J.}~\bibnamefont {Nipper}},
  \bibinfo {author} {\bibfnamefont {J.~B.}\ \bibnamefont {Balewski}}, \bibinfo
  {author} {\bibfnamefont {J.~P.}\ \bibnamefont {Shaffer}}, \bibinfo {author}
  {\bibfnamefont {R.}~\bibnamefont {Low}}, \bibinfo {author} {\bibfnamefont
  {T.}~\bibnamefont {Pfau}}, \bibinfo {author} {\bibfnamefont {W.}~\bibnamefont
  {Li}}, \bibinfo {author} {\bibfnamefont {J.}~\bibnamefont {Stanojevic}},
  \bibinfo {author} {\bibfnamefont {T.}~\bibnamefont {Pohl}}, \ and\ \bibinfo
  {author} {\bibfnamefont {J.~M.}\ \bibnamefont {Rost}},\ }\href@noop {}
  {\bibfield  {journal} {\bibinfo  {journal} {Phys. Rev. Lett.}\ } (\bibinfo
  {year} {Accepted for publication})}\BibitemShut {NoStop}%
\bibitem [{\citenamefont {Jaksch}\ \emph {et~al.}(2000)\citenamefont {Jaksch},
  \citenamefont {Cirac}, \citenamefont {Zoller}, \citenamefont {Rolston},\ and\
  \citenamefont {C\^{o}t\'{e}}}]{QIP1}%
  \BibitemOpen
  \bibfield  {author} {\bibinfo {author} {\bibfnamefont {D.}~\bibnamefont
  {Jaksch}}, \bibinfo {author} {\bibfnamefont {J.~I.}\ \bibnamefont {Cirac}},
  \bibinfo {author} {\bibfnamefont {P.}~\bibnamefont {Zoller}}, \bibinfo
  {author} {\bibfnamefont {S.~L.}\ \bibnamefont {Rolston}}, \ and\ \bibinfo
  {author} {\bibfnamefont {R.}~\bibnamefont {C\^{o}t\'{e}}},\ }\href@noop {}
  {\bibfield  {journal} {\bibinfo  {journal} {Phys. Rev. Lett.}\ }\textbf
  {\bibinfo {volume} {85}},\ \bibinfo {pages} {2208} (\bibinfo {year}
  {2000})}\BibitemShut {NoStop}%
\bibitem [{\citenamefont {Saffman}\ and\ \citenamefont
  {Walker}(2005)}]{FORTtheory}%
  \BibitemOpen
  \bibfield  {author} {\bibinfo {author} {\bibfnamefont {M.}~\bibnamefont
  {Saffman}}\ and\ \bibinfo {author} {\bibfnamefont {T.~G.}\ \bibnamefont
  {Walker}},\ }\href@noop {} {\bibfield  {journal} {\bibinfo  {journal} {Phys.
  Rev. A}\ }\textbf {\bibinfo {volume} {72}},\ \bibinfo {pages} {022347}
  (\bibinfo {year} {2005})}\BibitemShut {NoStop}%
\bibitem [{\citenamefont {Isenhower}\ \emph {et~al.}(2010)\citenamefont
  {Isenhower}, \citenamefont {Urban}, \citenamefont {Zhang}, \citenamefont
  {Gill}, \citenamefont {Henage}, \citenamefont {Johnson}, \citenamefont
  {Walker},\ and\ \citenamefont {Saffman}}]{QIP2}%
  \BibitemOpen
  \bibfield  {author} {\bibinfo {author} {\bibfnamefont {L.}~\bibnamefont
  {Isenhower}}, \bibinfo {author} {\bibfnamefont {E.}~\bibnamefont {Urban}},
  \bibinfo {author} {\bibfnamefont {X.~L.}\ \bibnamefont {Zhang}}, \bibinfo
  {author} {\bibfnamefont {A.~T.}\ \bibnamefont {Gill}}, \bibinfo {author}
  {\bibfnamefont {T.}~\bibnamefont {Henage}}, \bibinfo {author} {\bibfnamefont
  {T.~A.}\ \bibnamefont {Johnson}}, \bibinfo {author} {\bibfnamefont {T.~G.}\
  \bibnamefont {Walker}}, \ and\ \bibinfo {author} {\bibfnamefont
  {M.}~\bibnamefont {Saffman}},\ }\href@noop {} {\bibfield  {journal} {\bibinfo
   {journal} {Phys. Rev. Lett.}\ }\textbf {\bibinfo {volume} {104}},\ \bibinfo
  {pages} {10503} (\bibinfo {year} {2010})}\BibitemShut {NoStop}%
\bibitem [{\citenamefont {Gaëtan}\ \emph {et~al.}(2009)\citenamefont {Gaëtan},
  \citenamefont {Miroshnychenko}, \citenamefont {Wilk}, \citenamefont {Chotia},
  \citenamefont {Viteau}, \citenamefont {Comparat}, \citenamefont {Pillet},
  \citenamefont {Browaeys},\ and\ \citenamefont {Grangier}}]{QIP3}%
  \BibitemOpen
  \bibfield  {author} {\bibinfo {author} {\bibfnamefont {A.}~\bibnamefont
  {Gaëtan}}, \bibinfo {author} {\bibfnamefont {Y.}~\bibnamefont
  {Miroshnychenko}}, \bibinfo {author} {\bibfnamefont {T.}~\bibnamefont
  {Wilk}}, \bibinfo {author} {\bibfnamefont {A.}~\bibnamefont {Chotia}},
  \bibinfo {author} {\bibfnamefont {M.}~\bibnamefont {Viteau}}, \bibinfo
  {author} {\bibfnamefont {D.}~\bibnamefont {Comparat}}, \bibinfo {author}
  {\bibfnamefont {P.}~\bibnamefont {Pillet}}, \bibinfo {author} {\bibfnamefont
  {A.}~\bibnamefont {Browaeys}}, \ and\ \bibinfo {author} {\bibfnamefont
  {P.}~\bibnamefont {Grangier}},\ }\href@noop {} {\bibfield  {journal}
  {\bibinfo  {journal} {Nat. Phys.}\ }\textbf {\bibinfo {volume} {5}},\
  \bibinfo {pages} {115} (\bibinfo {year} {2009})}\BibitemShut {NoStop}%
\bibitem [{\citenamefont {Magalh$\tilde{\text{a}}$es}\ \emph
  {et~al.}(2000)\citenamefont {Magalh$\tilde{\text{a}}$es}, \citenamefont
  {de~Oliveira}, \citenamefont {Zanon}, \citenamefont {Bagnato},\ and\
  \citenamefont {Marcassa}}]{Marcassa1}%
  \BibitemOpen
  \bibfield  {author} {\bibinfo {author} {\bibfnamefont {K.~M.~F.}\
  \bibnamefont {Magalh$\tilde{\text{a}}$es}}, \bibinfo {author} {\bibfnamefont
  {A.~L.}\ \bibnamefont {de~Oliveira}}, \bibinfo {author} {\bibfnamefont {R.~A.
  D.~S.}\ \bibnamefont {Zanon}}, \bibinfo {author} {\bibfnamefont {V.~S.}\
  \bibnamefont {Bagnato}}, \ and\ \bibinfo {author} {\bibfnamefont {L.~G.}\
  \bibnamefont {Marcassa}},\ }\href@noop {} {\bibfield  {journal} {\bibinfo
  {journal} {Opt. Commun.}\ }\textbf {\bibinfo {volume} {184}},\ \bibinfo
  {pages} {385} (\bibinfo {year} {2000})}\BibitemShut {NoStop}%
\bibitem [{\citenamefont {Gabbanini}(2006)}]{PIcross}%
  \BibitemOpen
  \bibfield  {author} {\bibinfo {author} {\bibfnamefont {C.}~\bibnamefont
  {Gabbanini}},\ }\href@noop {} {\bibfield  {journal} {\bibinfo  {journal}
  {Spectrochim. Acta, Part B}\ }\textbf {\bibinfo {volume} {61}},\ \bibinfo
  {pages} {196} (\bibinfo {year} {2006})}\BibitemShut {NoStop}%
\bibitem [{\citenamefont {Tretyakov}\ \emph {et~al.}(2009)\citenamefont
  {Tretyakov}, \citenamefont {Beterov}, \citenamefont {Entin}, \citenamefont
  {Ryabtsev},\ and\ \citenamefont {Chapovsky}}]{Beterov}%
  \BibitemOpen
  \bibfield  {author} {\bibinfo {author} {\bibfnamefont {D.~B.}\ \bibnamefont
  {Tretyakov}}, \bibinfo {author} {\bibfnamefont {I.~I.}\ \bibnamefont
  {Beterov}}, \bibinfo {author} {\bibfnamefont {V.~M.}\ \bibnamefont {Entin}},
  \bibinfo {author} {\bibfnamefont {I.~I.}\ \bibnamefont {Ryabtsev}}, \ and\
  \bibinfo {author} {\bibfnamefont {P.~L.}\ \bibnamefont {Chapovsky}},\
  }\href@noop {} {\bibfield  {journal} {\bibinfo  {journal} {J. Exp. Theor.
  Phys.}\ }\textbf {\bibinfo {volume} {108}},\ \bibinfo {pages} {374} (\bibinfo
  {year} {2009})}\BibitemShut {NoStop}%
\bibitem [{\citenamefont {Feng}\ \emph {et~al.}(2009)\citenamefont {Feng},
  \citenamefont {Zhang}, \citenamefont {Zhao}, \citenamefont {Li},\ and\
  \citenamefont {Jia}}]{CsRyd}%
  \BibitemOpen
  \bibfield  {author} {\bibinfo {author} {\bibfnamefont {Z.-G.}\ \bibnamefont
  {Feng}}, \bibinfo {author} {\bibfnamefont {L.-J.}\ \bibnamefont {Zhang}},
  \bibinfo {author} {\bibfnamefont {J.-M.}\ \bibnamefont {Zhao}}, \bibinfo
  {author} {\bibfnamefont {C.-Y.}\ \bibnamefont {Li}}, \ and\ \bibinfo {author}
  {\bibfnamefont {S.-T.}\ \bibnamefont {Jia}},\ }\href@noop {} {\bibfield
  {journal} {\bibinfo  {journal} {J. Phys. B}\ }\textbf {\bibinfo {volume}
  {42}},\ \bibinfo {pages} {145303} (\bibinfo {year} {2009})}\BibitemShut
  {NoStop}%
\bibitem [{\citenamefont {Beterov}\ \emph {et~al.}(2009)\citenamefont
  {Beterov}, \citenamefont {Ryabtsev}, \citenamefont {Tretyakov},\ and\
  \citenamefont {Entin}}]{RydLifes}%
  \BibitemOpen
  \bibfield  {author} {\bibinfo {author} {\bibfnamefont {I.~I.}\ \bibnamefont
  {Beterov}}, \bibinfo {author} {\bibfnamefont {I.~I.}\ \bibnamefont
  {Ryabtsev}}, \bibinfo {author} {\bibfnamefont {D.~B.}\ \bibnamefont
  {Tretyakov}}, \ and\ \bibinfo {author} {\bibfnamefont {V.~M.}\ \bibnamefont
  {Entin}},\ }\href@noop {} {\bibfield  {journal} {\bibinfo  {journal} {Phys.
  Rev. A}\ }\textbf {\bibinfo {volume} {79}},\ \bibinfo {pages} {052504}
  (\bibinfo {year} {2009})}\BibitemShut {NoStop}%
\bibitem [{\citenamefont {Potvliege}\ and\ \citenamefont
  {Adams}(2006)}]{PIFORL}%
  \BibitemOpen
  \bibfield  {author} {\bibinfo {author} {\bibfnamefont {R.~M.}\ \bibnamefont
  {Potvliege}}\ and\ \bibinfo {author} {\bibfnamefont {C.~S.}\ \bibnamefont
  {Adams}},\ }\href@noop {} {\bibfield  {journal} {\bibinfo  {journal} {New J.
  Phys.}\ }\textbf {\bibinfo {volume} {8}},\ \bibinfo {pages} {145303}
  (\bibinfo {year} {2006})}\BibitemShut {NoStop}%
\bibitem [{\citenamefont {Sobelman}(1972)}]{Sobelman}%
  \BibitemOpen
  \bibfield  {author} {\bibinfo {author} {\bibfnamefont {I.~I.}\ \bibnamefont
  {Sobelman}},\ }\href@noop {} {\emph {\bibinfo {title} {Introduction to the
  Theory of Atomic Spectra}}},\ \bibinfo {edition} {1st}\ ed.\ (\bibinfo
  {publisher} {Elsevier},\ \bibinfo {year} {1972})\BibitemShut {NoStop}%
\bibitem [{\citenamefont {Tallant}\ \emph {et~al.}(2006)\citenamefont
  {Tallant}, \citenamefont {Overstreet}, \citenamefont {Schwettmann},\ and\
  \citenamefont {Shaffer}}]{RydTagging}%
  \BibitemOpen
  \bibfield  {author} {\bibinfo {author} {\bibfnamefont {J.}~\bibnamefont
  {Tallant}}, \bibinfo {author} {\bibfnamefont {K.~R.}\ \bibnamefont
  {Overstreet}}, \bibinfo {author} {\bibfnamefont {A.}~\bibnamefont
  {Schwettmann}}, \ and\ \bibinfo {author} {\bibfnamefont {J.~P.}\ \bibnamefont
  {Shaffer}},\ }\href@noop {} {\bibfield  {journal} {\bibinfo  {journal} {Phys.
  Rev. A}\ }\textbf {\bibinfo {volume} {74}},\ \bibinfo {pages} {023410}
  (\bibinfo {year} {2006})}\BibitemShut {NoStop}%
\bibitem [{\citenamefont {Kuppens}\ \emph {et~al.}(2000)\citenamefont
  {Kuppens}, \citenamefont {Corwin}, \citenamefont {Miller}, \citenamefont
  {Chupp},\ and\ \citenamefont {Wieman}}]{HFchange}%
  \BibitemOpen
  \bibfield  {author} {\bibinfo {author} {\bibfnamefont {S.~J.~M.}\
  \bibnamefont {Kuppens}}, \bibinfo {author} {\bibfnamefont {K.~L.}\
  \bibnamefont {Corwin}}, \bibinfo {author} {\bibfnamefont {K.~W.}\
  \bibnamefont {Miller}}, \bibinfo {author} {\bibfnamefont {T.~E.}\
  \bibnamefont {Chupp}}, \ and\ \bibinfo {author} {\bibfnamefont {C.~E.}\
  \bibnamefont {Wieman}},\ }\href@noop {} {\bibfield  {journal} {\bibinfo
  {journal} {Phys. Rev. A}\ }\textbf {\bibinfo {volume} {62}},\ \bibinfo
  {pages} {013406} (\bibinfo {year} {2000})}\BibitemShut {NoStop}%
\bibitem [{\citenamefont {Salvat}\ \emph {et~al.}(195)\citenamefont {Salvat},
  \citenamefont {Fern\'{a}ndez-Varea},\ and\ \citenamefont
  {W.~Williamson}}]{Radial}%
  \BibitemOpen
  \bibfield  {author} {\bibinfo {author} {\bibfnamefont {F.}~\bibnamefont
  {Salvat}}, \bibinfo {author} {\bibfnamefont {J.~M.}\ \bibnamefont
  {Fern\'{a}ndez-Varea}}, \ and\ \bibinfo {author} {\bibfnamefont
  {J.}~\bibnamefont {W.~Williamson}},\ }\href@noop {} {\bibfield  {journal}
  {\bibinfo  {journal} {Comput. Phys. Commun.}\ }\textbf {\bibinfo {volume}
  {90}},\ \bibinfo {pages} {151} (\bibinfo {year} {195})}\BibitemShut {NoStop}%
\bibitem [{\citenamefont {Marinescu}\ \emph {et~al.}(1994)\citenamefont
  {Marinescu}, \citenamefont {Sadeghpour},\ and\ \citenamefont
  {Dalgarno}}]{DispersionCoefs}%
  \BibitemOpen
  \bibfield  {author} {\bibinfo {author} {\bibfnamefont {M.}~\bibnamefont
  {Marinescu}}, \bibinfo {author} {\bibfnamefont {H.~R.}\ \bibnamefont
  {Sadeghpour}}, \ and\ \bibinfo {author} {\bibfnamefont {A.}~\bibnamefont
  {Dalgarno}},\ }\href@noop {} {\bibfield  {journal} {\bibinfo  {journal}
  {Phys. Rev. A}\ }\textbf {\bibinfo {volume} {49}},\ \bibinfo {pages} {982}
  (\bibinfo {year} {1994})}\BibitemShut {NoStop}%
\end{thebibliography}
%

\end{document}